\documentclass[aps,prl,twocolumn,superscriptaddress,showpacs,preprintnumbers,amsmath,amssymb]{revtex4}

\usepackage{graphicx}
\graphicspath{{fig/}}
\usepackage{dcolumn}
\usepackage{bm}
\usepackage{color}

\graphicspath{{figures/}}

\begin{document}

\preprint{APS/123-QED}

\title{A Statistical Model of  Aggregates Fragmentation}

\author{F. Spahn}
\affiliation{Institute of Physics, University of Potsdam,
  Am Neuen Palais 10, 14469 Potsdam, Germany}
\author{E. V. Neto}
 \affiliation{Grupo de Din\^{a}mica Orbital e Planetologia, UNESP, S\~ao Paulo, Brazil
 }
\author{A. H. F. Guimar\~{a}es}
\affiliation{Institute of Physics, University of Potsdam,
  Am Neuen Palais 10, 14469 Potsdam, Germany}
\author{A. N. Gorban }%
\affiliation{%
Department of Mathematics, University of Leicester, Leicester LE1 7RH, United Kingdom
}%
\author{N. V. Brilliantov}%
\affiliation{%
Department of Mathematics, University of Leicester, Leicester LE1 7RH, United Kingdom
}%

\date{\today}

\begin{abstract}
A statistical model of fragmentation of aggregates is proposed, based on the stochastic propagation of
cracks through the body. The propagation rules are formulated on a lattice and mimic two important
features of the process -- a crack  moves against the stress gradient and its energy depletes as it
grows. We perform numerical simulations  of the model for two-dimensional lattice and reveal that the
mass  distribution for small and intermediate-size  fragments obeys a power-law, $F(m) \propto
m^{-3/2}$, in agreement with experimental observations.  We develop an analytical theory which explains
the detected power-law   and demonstrate that the overall fragment mass distribution in our model agrees
qualitatively with that, observed in experiments.

\end{abstract}

\maketitle


{\it Introduction.}~ Fragmentation processes are ubiquitous in nature and play an important role in many
industrial processes. Numerous examples range from manufacturing comminution to collision of cosmic
bodies in space. Hence,  much effort has been devoted to comprehend the nature of fragmentation in
different branches of research -- geophysics \citep{Grady1980,Turcotte1997}, astrophysics
\citep{Michel2003,Nakamura2008}, engineering \citep{Thornton1996} and  material- \citep{Lankford1991} or
military science \citep{Mott1943,Grady2006}.

An intriguing common property of fragmentation is a power law mass-distribution of the fragments which
seems to be independent of the spatial scale or nature of the  parent bodies: When heavy ions collide
with a target,  or asteroids suffer a high-speed  impact, both produce a power law size (or charge)
distribution of debris. This law has been reported in numerous experimental studies, e.g.
\cite{Herrmann1990,Nakamura-e-Fujiwara-1991,Giblin-etal-1998,Arakawa1999,Ryan-2000,Herrmann-etal-2006,Kun-etal-2006,
Guetler2010,Herrmann2010}.

The theoretical description of fragmentation is  developing along two different lines: one, based on
continuum mechanics, another -- on the general statistical methods,
e.g.~\cite{Herrmann1990,Astrom2004,Grady2009}. Finite element analysis (FEM) \cite{Chang-etal-2002b},
numerical simulations of agglomerate collisions with smooth particle hydrodynamics (SMH)
\cite{Benz-e-Asphaug-1994}, or discrete element method (DEM) \cite{Herrmann-etal-2006,Herrmann2010} --
these are the current tools to treat continuum mechanical problem of colliding material bodies.
Alternatively, statistical methods, as random walk \cite{Bouchaud-etal-1993} or stochastic simulation of
crack-tip trajectories \cite{Galybin-e-Dyskin-2004} were used to model fractal crack formation and
propagation. Furthermore, there were attempts to tackle the problem analytically. In particular, a
semi-empirical approach to the physics of catastrophic  breakup of cosmic bodies has been developed
\cite{Paolicchi-etal-1996}; similarly, a micro-structural approach has been adopted
\cite{Chang-etal-2002a} to model  the fracture generation and propagation in concrete.

The observed universality of the fragmentation law for the objects, drastically different in material
properties and dimension, most probably implies a  common physical principle  inherent to all
fragmentation processes~\citep{Kun1999}. Moreover, it is reasonable to assume that this principle is of
statistical nature. This motivates the analysis of a model, which being very simple on the microscopic
level, reflects the most prominent features of the process and adequately represents its statistical
properties. In the present Letter we propose a novel statistical model for fragmentation of  aggregates
-- macroscopic bodies,  comprised of  a large amount of smaller macroscopic constituents. The
problem of aggregate fragmentation arises in many areas of science and technology, in particular in
planetary science, where the availability of an adequate fragmentation model is crucial for
understanding of the formation and evolution of planetary rings,
e.g.~\cite{Longaretti1996,Spahn-etal-2004,Sremcevic2007}. We show that our model, based on a few very
simple physical rules reproduces the main aspects of the fragmentation process and gives the power-law
distribution for the debris size. Moreover, it qualitatively agrees with the experimental data. We
perform  numerical simulations and develop a simple analytical theory, which explains the observed
power-law distribution for the fragments.

\begin{figure}[thb]
\begin{center}
\includegraphics[scale=0.8]{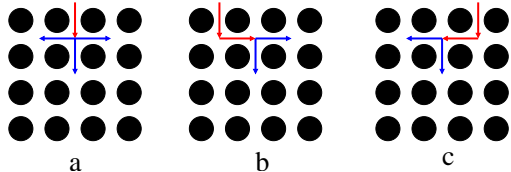}
\caption{(Color on line). A model aggregate is composed of $l \times l$ identical particles on a square
lattice. To break a bond between particles the energy $E_{\rm b}$ is needed. A crack propagates either
away from the loaded surface (vertically-down on the plots), against the stress gradient, or laterally,
but never along the gradient. All possible directions have equal probability. In the case (a) there
exist three allowed directions, each with the probability $p_3=1/3$: one  away from the surface and two
lateral directions. In the cases (b) and (c) only two directions with the probability $p_2=1/2$ are
possible, since crack can not move back or "uphill" the load. \label{fig:model} }
\end{center}
\end{figure}

{\it Formulation of the model.} We consider aggregates, which are not too large to neglect the
gravitation forces and assume that particles  are kept together by the adhesive bonds. These are
significantly weaker than the forces attributed to chemical bonds, hence the formation of cracks occurs
only along the contact lines between the particles. We assume that all particles are of the same size
and material. Then  to break any  adhesive contact,  the amount of energy, equal to $E_{\rm b}$, is
required; to destroy simultaneously $n$ bonds one needs $n$-times larger energy, $nE_{\rm b}$. The
quantity $E_{\rm b}$ depends on the particles size, their surface tension and material parameters, e.g.
\cite{Brilliantov2007}. To simplify further the problem, we consider its two-dimensional version and
 assume that the spherical constituents form a square lattice, Fig.~\ref{fig:model}.

In a collision between aggregates, which causes their fragmentation, the  energy of the relative motion
$E_{\rm coll}$ is transformed into the surface energy of broken bonds  and the kinetic energy of debris.
We will ignore the latter part of the energy and explicitly consider the former one.
Therefore, the problem of fragmentation in our model is, essentially, reduced to the problem of
distribution of the energy $E_{\rm coll}$ among the broken adhesive bonds. We assume the bonds being
destroyed due the crack's propagation. Namely, we adopt following rules for the crack
growth:~(i)~A crack originates on the surface, on the impact site, where the maximum stress is expected.
It propagates (in average) away from the surface against the stress gradient, that is, "downhill" the
load, and never "uphill". (ii)~Each time-step the crack elongates by a single neighboring bond,
consuming the energy $E_{\rm b}$, while the crack tip performs a random walk with the direction randomly
chosen among two or three possible ones, as explained in Fig.~\ref{fig:model}. (iii)~The propagation of
a certain crack terminates if its tip arrives at the surface, or meets another crack, mimicking a crack
bifurcation. When a crack terminates and the rest-energy allows for further bond-breaking, a new crack
is initiated randomly on the surface of the aggregate, or bifurcates from another crack. (iv)~A
mechanism limiting a total crack energy is introduced.  This is because crack generation and propagation
needs a sufficient load gradient which decreases with distance from the impact site. Thus, we
assume that if the energy assigned to a crack, chosen randomly between zero and a maximum $E_{\rm
cross}$, is exhausted, the  crack stops and another one is created instead. We called this
{\it "crossing factor"}. (v)~The breaking process continues as long as the remaining impact energy,
$E_{\rm rest} = E_{\rm coll}-n E_{\rm b} \,$ ($n$ -- current number of broken bonds) is sufficient to
break further bonds. It terminates whenever the total impact energy $E_{\rm coll}$ is dissipated in
cracks.

\begin{figure}[htbp]
{   \centerline{\includegraphics[width=0.45\columnwidth]{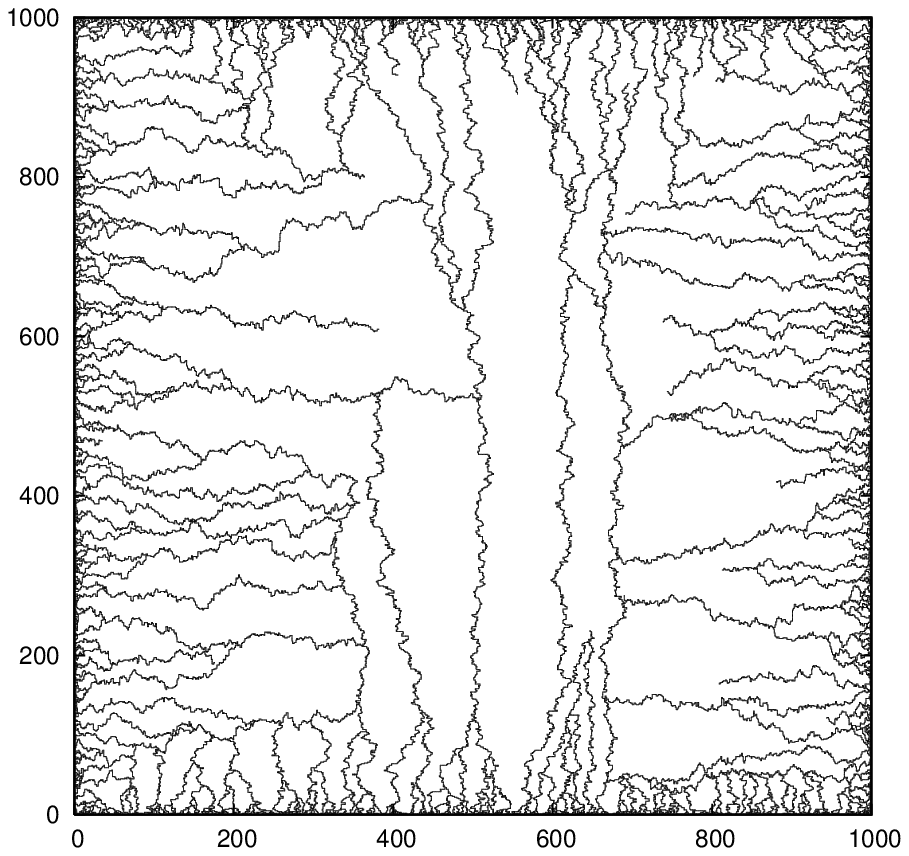}~
                \includegraphics[width=0.45\columnwidth]{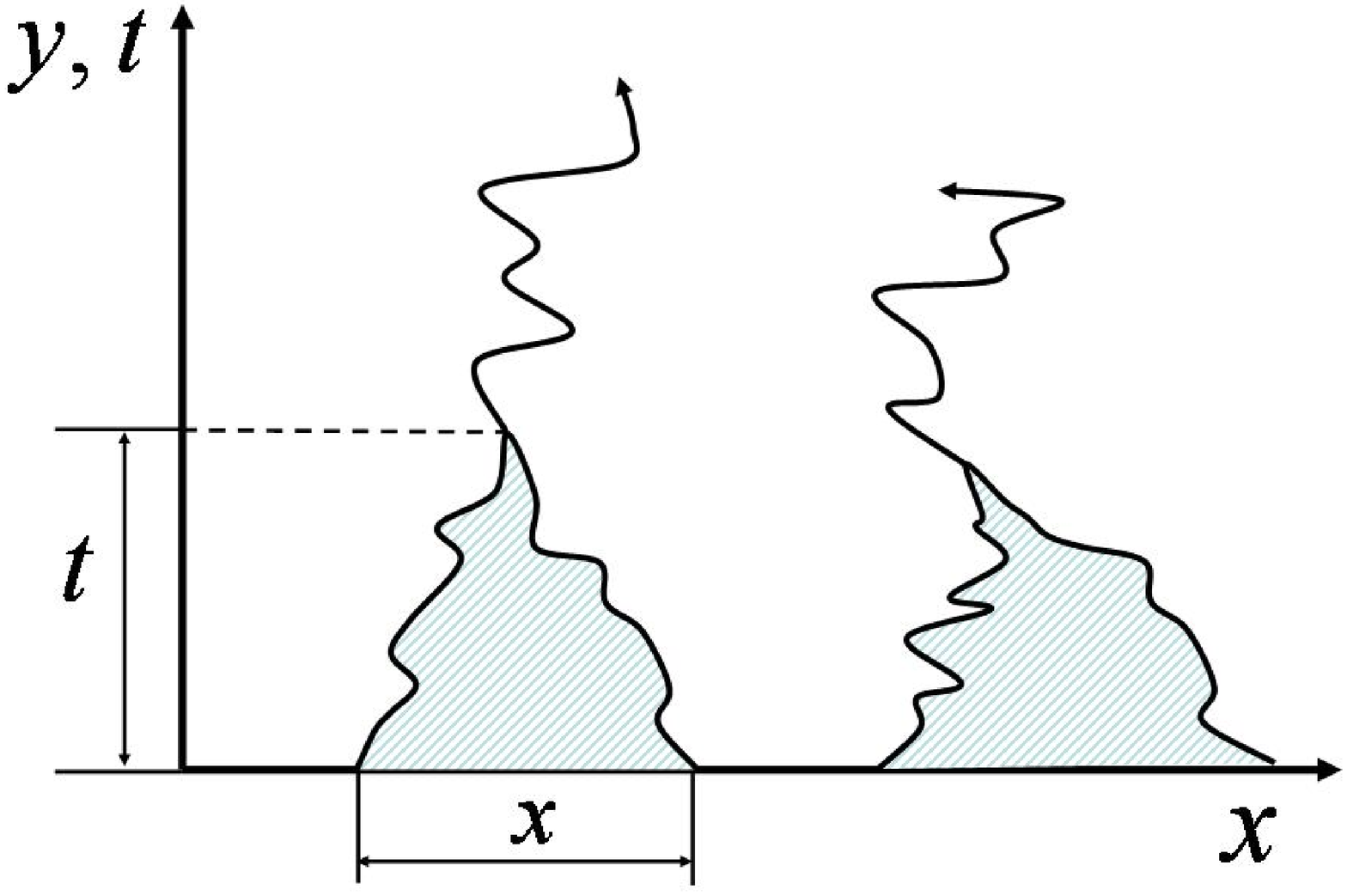}}
}
  \caption{Left panel:~ A typical fragmentation pattern. The aggregate, composed of one million particles on the
$1\,000\times1\,000$ lattice,  is broken in 3958 pieces. The total breakage energy is $E_{\rm
coll}=82\,010 E_{\rm b}$, the crossing-factor is  $E_{\rm cross}=3\,000~E_{\rm b}$. Right panel: The
fragmentation model may be mapped on the one-dimensional model of diffusing and annihilating particles.
The areas under the particle's trajectories correspond to the fragment areas of the primary
fragmentation model.}
  \label{fig:pattern_model}
\end{figure}

\vspace*{-0.6cm} In Fig.~\ref{fig:pattern_model} a typical fragmentation pattern, obtained with the use
of the above fragmentation rules, is shown. Next, we analyze the statistical properties of debris.

{\it Fragment mass distribution.} We define a fragment as a collection of constituent particles connected
by adhesive bonds where a single particle consitutes a mass unit. Hence, for a
fragment of  $m$ particles  we assign the mass $m$. We have performed numerical experiments, controlling
the size of the aggregate, the total energy spent in the process and the crossing factor. To improve the
fragments' statistics we perform many runs with the same set of parameters.

\begin{figure}[thb]
\begin{center}\includegraphics[scale=0.45]{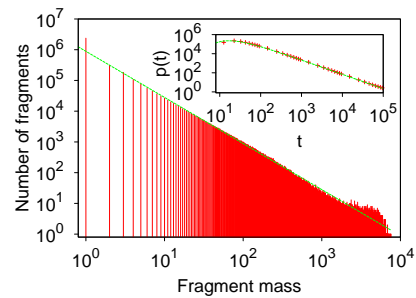}
  \caption{(Color online). Fragment mass distribution for a $100\times100$ lattice
  matrix with total energy of $4\,000 ~E_{\rm b}$ and a crossing factor of $200~E_{\rm b}$.
   The data have been collected for   $10\,000$ runs.
   The line represents  a power law mass-distribution with an exponent $\simeq - 3/2$. The inset
   shows the distribution function $P(x,t) \propto t^{-3/2}$, which gives the probability that two particles,
   initially separated by  distance $x$, meet at time $t$; it illustrates the affinity of the
   two models (see the text). Points - numerical data for $x=20$, line - theory.
\label{fig:fragcem}}
\end{center}
\end{figure}

Fig.~\ref{fig:fragcem} shows the fragment mass-distribution. The most part of the distribution, except
for very large fragments, may be accurately fitted by the power-law
\begin{equation}
F(m) \propto m^{- \alpha} \, . \label{eq:1}
\end{equation}
Interestingly, the exponent $\alpha \simeq 1.5$ is the same as in 2D egg-shell crushing experiments
\citep{Herrmann-etal-2006}.
\newcommand{\scl}{0.7}

 {\it Analytical model.}  The observed power-law fragment size distribution
may be obtained analytically, if we notice that the proposed fragmentation
model may be mapped onto the  one-dimensional model of diffusing and annihilating particles, where the
particles correspond to  the tips of the cracks. Indeed, the direction normal to the surface, that
is, the direction of the most rapid decay of the stress, say axis $y$ (vertical downward in Fig.
\ref{fig:model}),  may be mapped onto the time axis, since the reverse motion along this axis is
forbidden. Then the location of the crack tips along the lateral direction, say along axis $x$
(horizontal in Fig. \ref{fig:model}) corresponds to the location of fictitious "particles" on the line.
One step along $y$ axis corresponds to one time step, while one step along $x$ axis corresponds to a
"particle" jump on the line (without loss of generality we can assume unit time and space steps).

It is easy to see that the fragmentation model with the rules illustrated in Fig. \ref{fig:model} is
equivalent to the following model: (i)~Each particle on a line at each time step can remain at the same
site with the probability $p_0=1/3$, move one site left or right with the probability $p_1=1/6$, two
sites left or right with  $p_2=1/12$, etc., $k$ sites left or right with the probability $p_k=1/(3\cdot
2^k)$; note that $p_0+2\sum_k p_k=1$. (ii)~When two particles meet, one of them  annihilates while the
other one continues to move with the same rules.  (iii)~The fragment sizes of the primary fragmentation
model is equal to the area confined by the trajectories of the particles and the axis $x$,
Fig.~\ref{fig:pattern_model}.

First we show that the "particles" perform one-dimensional diffusion motion. Indeed, due to the lack of
memory, each time step does not depend on  the previous one (note that the original problem does have a
memory with respect to previous steps). Therefore, the mean square displacement $ < \left[ \Delta x(t)
\right]^2 >$ is a sum of these quantities for each step, $\left< \Delta ^2 \right>$, where
$$
\left<  \Delta^2 \right> =\sum_{l=0}^{\infty} 2\, p_ll^2 \, = \,  \sum_{l=0}^{\infty} \frac{2\, l^2}{3 \cdot
2^l}=4\, ,
$$
that is, $ \left< \left[ \Delta x(t) \right]^2 \right> = 4t=2Dt$.  Hence, we have a diffusive motion
with the diffusion coefficient $D=2$.

Now we compute the distribution of the fragments' areas. These correspond to the areas of the figures on
the $x-t$ plane confined by the $x$-axis  and the diffusive trajectories of pairs of particles, which
start to move at  $t=0$, and meet at $t >0$; the initial separation of the particles is $x$. The area of
the figures, that is, the fragment mass scales as $m \propto t x$. Hence, the mass distribution reads:
\begin{equation}
\label{eq:2} F(m) \simeq  \int dx \int dt\delta (m - \gamma t x) \varkappa e^{-\varkappa \, x}
P(x,t)\, ,
\end{equation}
where the coefficient $\gamma$ accounts for the surface density and a geometric factor (1/2
for triangles), which relates  a fragment area and its dimensions $x$ and $t$. The coefficient
$\varkappa$ is the line density of particles on the $x$ axis at the starting time $t=0$ (i.e. the
density of crack tips which start at $y=0$); the random initial distribution of particles (crack tips)
implies the Poisson distribution for the initial inter-particle distance, $\varkappa \exp(- \varkappa
x)$.  \, $P(x,t)$ gives the probability of two diffusing particles, initially
separated by distance $x$, to meet at time $t$; where one of the particles annihilates. $P(x,t)$
may be written as $P(x,t)=\frac{d}{dt}P_{\rm surv}(x,t)$, where
\begin{equation}
\label{eq:3} P_{\rm surv}(x,t) \!=\!\int_0^{\infty}\!\!\! \!\! \frac{1}{\sqrt{8\pi Dt}} \left(
e^{-(y-x)^2/8Dt} - e^{-(y+x)^2/8Dt} \right) dy
\end{equation}
is the survival probability, i.e. that both particles have not annihilated before
time $t$, provided they started to diffuse at $t=0$ separated by the distance $x$. It is obtained from
the solution of the diffusion problem with the adsorbing boundary condition for the relative motion of
two particles, with the double diffusion coefficient $2D$ (see e.g. \cite{Krap2010}). Indeed the
integrand in Eq.~(\ref{eq:3}) gives the probability distribution of the
inter-particle distance $y$ at time $t$, if the initial separation was $x$, provided this probability is
identically zero for $y=0$ (the annihilation condition). Integrating this probability over all distances
$y>0$ gives the survival probability; differentiation of $P_{\rm surv}(x,t)$ with respect to time yields
the probability that the particles meet  exactly at time $t$. Substituting $P(x,t)=x \, \exp(-x^2/8Dt) /\sqrt{32 Dt^3}$, obtained from Eq.~(\ref{eq:3}) into Eq.~(\ref{eq:2}), we arrive at the expansion,
\begin{equation}
\label{eq:4} F(m) = A_0 m^{-3/2} + A_1 m^{-5/2} + \ldots\, ,
\end{equation}
where $A_0=\gamma^{1/2}\Gamma(5/2)/8\varkappa^{3/2}$ and generally,
$A_l=(-1)^l\Gamma(3l+5/2)(\gamma/\varkappa)^{3l+1/2}/(8\varkappa)(4 \gamma)^{2l}l!$ is the coefficient
at $m^{-(3/2+l)}$. Hence, Eq.~(\ref{eq:4}) explains the power-law distribution (\ref{eq:1}) obtained
numerically. The above expansion holds,  however,  not  for very small masses,
to keep the continuum diffusion approach valid,  and not for  very large masses, to ignore the effects
of energy depletion and finite size effects of the fragmenting pattern.

\begin{figure}[thb]
\begin{center}
\includegraphics[scale=0.85,angle=0]{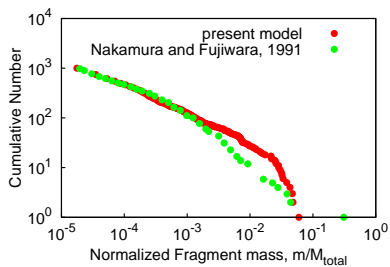}
\caption{(Color online). Results of our model (red circles) compared with the experimental data (green
  circles) of \citet{Nakamura-e-Fujiwara-1991}. In simulations we used a $80 \times 80$ matrix, the
total energy of $E_{\rm coll}=1500~E_{\rm b}$, the crossing-factor of  $E_{\rm cross}=100~E_{\rm b}$;
the number of runs is 10.  \label{fig:comp01}}
\end{center}
\end{figure}

{\it Comparison with experiments.} So far we considered the size distribution of fragments which mass is
much smaller than the initial mass of the parent body. In this limit the numerically detected and
explained power-law distribution coincides with the experimental one \cite{Herrmann-etal-2006}.
\begin{figure}[thb]
\begin{center}
\includegraphics[scale=0.85,angle=0]{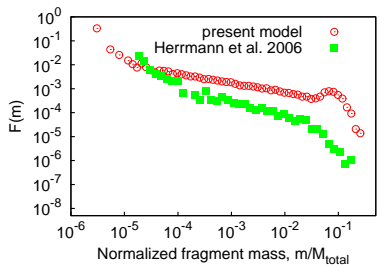}
\caption{(Color online). Results of our model (red circles) compared with the experimental data (green
squares) of \citet{Herrmann-etal-2006}. In simulations we used a $600 \times 600$ matrix, the total
energy of $E_{\rm coll}=40\,000~E_{\rm b}$, the crossing-factor of  $E_{\rm cross}=3\,000~E_{\rm b}$ and
the number of runs is 200. Note the lack of any fitting or scaling of our data. \label{fig:comp02}}
\end{center}
\end{figure}
Surprisingly, our model can qualitatively and sometimes even semi-quantitatively reproduce the whole
size distribution of fragments in the impact experiments.
This is illustrated in Fig.~\ref{fig:comp02} for the experiments of exploding egg-shells
\cite{Herrmann-etal-2006}
and in Fig.~\ref{fig:comp01} for the
colliding basalt spheres \cite{Nakamura-e-Fujiwara-1991}. For the former case our model demonstrates an
overall qualitatively correct  behavior for $F(m)$,  while for the latter case one can achieve a
quantitative agreement with the most part of the experimental fragment size distribution. It is worth noting
 that the agreement is obtained without any fitting or scaling of our data. Moreover, the fact that our
simple two-dimensional model can describe the fragmentation statistics for  three-dimensional bodies
 indicates, that the main ingredients of our model - the diffusive propagation of cracks
against the stress gradient and the depletion of energy with the cracks' growth - are indeed the basic
features of a fragmentation process.

 {\it In conclusion} we suggest a simple fragmentation model, based on a lattice random
walk of a crack with propagation rules which mimic a real fragmentation process: A crack moves against
(or laterally to) the stress gradient and the energy of the crack exhausts along its growth. We
performed numerical simulations of the model and observed that the mass distribution of small and
intermediate fragments obeys a power-law. The exponent of the power law, equal to $-3/2$, agrees  with
the reported experimental value. We develop an analytical theory which explains the nature of the
power-law in the fragmentation statistics and demonstrate that the proposed model gives a qualitative
description for the overall fragment size distribution observed in experiments.


\begin{thebibliography}{31}
\expandafter\ifx\csname natexlab\endcsname\relax\def\natexlab#1{#1}\fi
\expandafter\ifx\csname bibnamefont\endcsname\relax
  \def\bibnamefont#1{#1}\fi
\expandafter\ifx\csname bibfnamefont\endcsname\relax
  \def\bibfnamefont#1{#1}\fi
\expandafter\ifx\csname citenamefont\endcsname\relax
  \def\citenamefont#1{#1}\fi
\expandafter\ifx\csname url\endcsname\relax
  \def\url#1{\texttt{#1}}\fi
\expandafter\ifx\csname urlprefix\endcsname\relax\def\urlprefix{URL }\fi
\providecommand{\bibinfo}[2]{#2}
\providecommand{\eprint}[2][]{\url{#2}}

\bibitem[{\citenamefont{{Grady} and {Lipkin}}(1980)}]{Grady1980}
\bibinfo{author}{\bibfnamefont{D.~E.} \bibnamefont{{Grady}}} \bibnamefont{and}
  \bibinfo{author}{\bibfnamefont{J.}~\bibnamefont{{Lipkin}}},
  \bibinfo{journal}{Geophys. Res. Lett.} \textbf{\bibinfo{volume}{7}},
  \bibinfo{pages}{255} (\bibinfo{year}{1980}).

\bibitem[{\citenamefont{{Turcotte}}(1997)}]{Turcotte1997}
\bibinfo{author}{\bibfnamefont{D.~L.} \bibnamefont{{Turcotte}}},
  \emph{\bibinfo{title}{Fractals and Chaos in Geology and Geophysics}}
  (\bibinfo{publisher}{Cambridge}, \bibinfo{year}{1997}).

\bibitem[{\citenamefont{{Michel} et~al.}(2003)\citenamefont{{Michel}, {Benz},
  and {Richardson}}}]{Michel2003}
\bibinfo{author}{\bibfnamefont{P.}~\bibnamefont{{Michel}}},
  \bibinfo{author}{\bibfnamefont{W.}~\bibnamefont{{Benz}}}, \bibnamefont{and}
  \bibinfo{author}{\bibfnamefont{D.~C.} \bibnamefont{{Richardson}}},
  \bibinfo{journal}{\nat} \textbf{\bibinfo{volume}{421}}, \bibinfo{pages}{608}
  (\bibinfo{year}{2003}).

\bibitem[{\citenamefont{{Nakamura} et~al.}(2008)\citenamefont{{Nakamura},
  {Michikami}, {Hirata}, {Fujiwara}, {Nakamura}, {Ishiguro}, {Miyamoto},
  {Demura}, {Hiraoka}, {Honda} et~al.}}]{Nakamura2008}
\bibinfo{author}{\bibfnamefont{A.~M.} \bibnamefont{{Nakamura}}},
  \bibinfo{author}{\bibfnamefont{T.}~\bibnamefont{{Michikami}}},
  \bibinfo{author}{\bibfnamefont{N.}~\bibnamefont{{Hirata}}},
  \bibinfo{author}{\bibfnamefont{A.}~\bibnamefont{{Fujiwara}}},
  \bibinfo{author}{\bibfnamefont{R.}~\bibnamefont{{Nakamura}}},
  \bibinfo{author}{\bibfnamefont{M.}~\bibnamefont{{Ishiguro}}},
  \bibinfo{author}{\bibfnamefont{H.}~\bibnamefont{{Miyamoto}}},
  \bibinfo{author}{\bibfnamefont{H.}~\bibnamefont{{Demura}}},
  \bibinfo{author}{\bibfnamefont{K.}~\bibnamefont{{Hiraoka}}},
  \bibinfo{author}{\bibfnamefont{T.}~\bibnamefont{{Honda}}},
  \bibnamefont{et~al.}, \bibinfo{journal}{Earth, Planets, and Space}
  \textbf{\bibinfo{volume}{60}}, \bibinfo{pages}{7} (\bibinfo{year}{2008}).

\bibitem[{\citenamefont{{Thornton} et~al.}(1996)\citenamefont{{Thornton},
  {Yin}, and {Adams}}}]{Thornton1996}
\bibinfo{author}{\bibfnamefont{C.}~\bibnamefont{{Thornton}}},
  \bibinfo{author}{\bibfnamefont{K.~K.} \bibnamefont{{Yin}}}, \bibnamefont{and}
  \bibinfo{author}{\bibfnamefont{M.~J.} \bibnamefont{{Adams}}},
  \bibinfo{journal}{J. Phys. D: Appl. Phys.} \textbf{\bibinfo{volume}{29}},
  \bibinfo{pages}{424} (\bibinfo{year}{1996}).

\bibitem[{\citenamefont{{Lankford} and {Blanchard}}(1991)}]{Lankford1991}
\bibinfo{author}{\bibfnamefont{J.}~\bibnamefont{{Lankford}}} \bibnamefont{and}
  \bibinfo{author}{\bibfnamefont{C.~R.} \bibnamefont{{Blanchard}}},
  \bibinfo{journal}{J. Mat. Sci.} \textbf{\bibinfo{volume}{26}},
  \bibinfo{pages}{3067} (\bibinfo{year}{1991}).

\bibitem[{\citenamefont{{Mott} and {Linfoot}}(1943)}]{Mott1943}
\bibinfo{author}{\bibfnamefont{N.~F.} \bibnamefont{{Mott}}} \bibnamefont{and}
  \bibinfo{author}{\bibfnamefont{E.~H.} \bibnamefont{{Linfoot}}},
  \bibinfo{type}{Tech. Rep.}, \bibinfo{institution}{Ministry of Supply}
  (\bibinfo{year}{1943}).

\bibitem[{\citenamefont{Grady}(2006)}]{Grady2006}
\bibinfo{author}{\bibfnamefont{D.~E.} \bibnamefont{Grady}},
  \emph{\bibinfo{title}{{Fragmentation of rings and shells: The legacy of N. F.
  Mott}}} (\bibinfo{publisher}{Springer}, \bibinfo{year}{2006}).

\bibitem[{\citenamefont{{Herrmann} and {Roux}}(1990)}]{Herrmann1990}
\bibinfo{author}{\bibfnamefont{H.~J.} \bibnamefont{{Herrmann}}}
  \bibnamefont{and} \bibinfo{author}{\bibfnamefont{S.}~\bibnamefont{{Roux}}},
  \emph{\bibinfo{title}{Statistical Models for the Fracture of Disordered
  Media}} (\bibinfo{publisher}{Elsevier}, \bibinfo{year}{1990}).

\bibitem[{\citenamefont{{Nakamura} and
  {Fujiwara}}(1991)}]{Nakamura-e-Fujiwara-1991}
\bibinfo{author}{\bibfnamefont{A.}~\bibnamefont{{Nakamura}}} \bibnamefont{and}
  \bibinfo{author}{\bibfnamefont{A.}~\bibnamefont{{Fujiwara}}},
  \bibinfo{journal}{Icarus} \textbf{\bibinfo{volume}{92}}, \bibinfo{pages}{132}
  (\bibinfo{year}{1991}).

\bibitem[{\citenamefont{{Giblin} et~al.}(1998)\citenamefont{{Giblin},
  {Martelli}, {Farinella}, {Paolicchi}, {di Martino}, and
  {Smith}}}]{Giblin-etal-1998}
\bibinfo{author}{\bibfnamefont{I.}~\bibnamefont{{Giblin}}},
  \bibinfo{author}{\bibfnamefont{G.}~\bibnamefont{{Martelli}}},
  \bibinfo{author}{\bibfnamefont{P.}~\bibnamefont{{Farinella}}},
  \bibinfo{author}{\bibfnamefont{P.}~\bibnamefont{{Paolicchi}}},
  \bibinfo{author}{\bibfnamefont{M.}~\bibnamefont{{di Martino}}},
  \bibnamefont{and} \bibinfo{author}{\bibfnamefont{P.~N.}
  \bibnamefont{{Smith}}}, \bibinfo{journal}{Icarus}
  \textbf{\bibinfo{volume}{134}}, \bibinfo{pages}{77} (\bibinfo{year}{1998}).

\bibitem[{\citenamefont{Arakawa}(1999)}]{Arakawa1999}
\bibinfo{author}{\bibfnamefont{M.}~\bibnamefont{Arakawa}},
  \bibinfo{journal}{Icarus} \textbf{\bibinfo{volume}{142}},
  \bibinfo{pages}{34–45} (\bibinfo{year}{1999}).

\bibitem[{\citenamefont{{Ryan}}(2000)}]{Ryan-2000}
\bibinfo{author}{\bibfnamefont{E.~V.} \bibnamefont{{Ryan}}},
  \bibinfo{journal}{Annu. Rev. Earth Planet. Sci.}
  \textbf{\bibinfo{volume}{28}}, \bibinfo{pages}{367} (\bibinfo{year}{2000}).

\bibitem[{\citenamefont{{Herrmann} et~al.}(2006)\citenamefont{{Herrmann},
  {Wittel}, and {Kun}}}]{Herrmann-etal-2006}
\bibinfo{author}{\bibfnamefont{H.~J.} \bibnamefont{{Herrmann}}},
  \bibinfo{author}{\bibfnamefont{F.~K.} \bibnamefont{{Wittel}}},
  \bibnamefont{and} \bibinfo{author}{\bibfnamefont{F.}~\bibnamefont{{Kun}}},
  \bibinfo{journal}{Physica A} \textbf{\bibinfo{volume}{371}},
  \bibinfo{pages}{59} (\bibinfo{year}{2006}).

\bibitem[{\citenamefont{{Kun} et~al.}(2006)\citenamefont{{Kun}, {Wittel},
  {Herrmann}, {Kr\"{o}plin}, and {Maloy}}}]{Kun-etal-2006}
\bibinfo{author}{\bibfnamefont{F.}~\bibnamefont{{Kun}}},
  \bibinfo{author}{\bibfnamefont{F.~K.} \bibnamefont{{Wittel}}},
  \bibinfo{author}{\bibfnamefont{H.~J.} \bibnamefont{{Herrmann}}},
  \bibinfo{author}{\bibfnamefont{B.~H.} \bibnamefont{{Kr\"{o}plin}}},
  \bibnamefont{and} \bibinfo{author}{\bibfnamefont{K.~J.}
  \bibnamefont{{Maloy}}}, \bibinfo{journal}{\prl}
  \textbf{\bibinfo{volume}{96}}, \bibinfo{pages}{025504}
  (\bibinfo{year}{2006}).

\bibitem[{\citenamefont{Guettler et~al.}(2010)\citenamefont{Guettler, Blum,
  Zsom, Ormel, and Dullemond}}]{Guetler2010}
\bibinfo{author}{\bibfnamefont{C.}~\bibnamefont{Guettler}},
  \bibinfo{author}{\bibfnamefont{J.}~\bibnamefont{Blum}},
  \bibinfo{author}{\bibfnamefont{A.}~\bibnamefont{Zsom}},
  \bibinfo{author}{\bibfnamefont{C.~W.} \bibnamefont{Ormel}}, \bibnamefont{and}
  \bibinfo{author}{\bibfnamefont{C.~P.} \bibnamefont{Dullemond}},
  \bibinfo{journal}{Astronomy and Astrophysics} \textbf{\bibinfo{volume}{A56}},
  \bibinfo{pages}{513} (\bibinfo{year}{2010}).

\bibitem[{\citenamefont{Timar et~al.}(2010)\citenamefont{Timar, Blomer, Kun,
  and Herrmann}}]{Herrmann2010}
\bibinfo{author}{\bibfnamefont{G.}~\bibnamefont{Timar}},
  \bibinfo{author}{\bibfnamefont{J.}~\bibnamefont{Blomer}},
  \bibinfo{author}{\bibfnamefont{F.}~\bibnamefont{Kun}}, \bibnamefont{and}
  \bibinfo{author}{\bibfnamefont{H.~J.} \bibnamefont{Herrmann}},
  \bibinfo{journal}{Phys. Rev. Lett.} \textbf{\bibinfo{volume}{104}},
  \bibinfo{pages}{095502} (\bibinfo{year}{2010}).

\bibitem[{\citenamefont{Astrom et~al.}(2004)\citenamefont{Astrom, Ouchterlony,
  and Linna}}]{Astrom2004}
\bibinfo{author}{\bibfnamefont{J.~A.} \bibnamefont{Astrom}},
  \bibinfo{author}{\bibfnamefont{F.}~\bibnamefont{Ouchterlony}},
  \bibnamefont{and} \bibinfo{author}{\bibfnamefont{J.}~\bibnamefont{Linna},
  \bibfnamefont{R.~P.~andTimonen}}, \bibinfo{journal}{Phys. Rev. Lett.}
  \textbf{\bibinfo{volume}{92}}, \bibinfo{pages}{245506}
  (\bibinfo{year}{2004}).

\bibitem[{\citenamefont{Grady}(2009)}]{Grady2009}
\bibinfo{author}{\bibfnamefont{D.}~\bibnamefont{Grady}}, \bibinfo{journal}{Int.
  J. Fract.} pp. \bibinfo{pages}{1--15} (\bibinfo{year}{2009}).

\bibitem[{\citenamefont{{Chang}
  et~al.}(2002{\natexlab{a}})\citenamefont{{Chang}, {Wang}, {Sluys}, and {van
  Mier}}}]{Chang-etal-2002b}
\bibinfo{author}{\bibfnamefont{C.~S.} \bibnamefont{{Chang}}},
  \bibinfo{author}{\bibfnamefont{T.~K.} \bibnamefont{{Wang}}},
  \bibinfo{author}{\bibfnamefont{L.~J.} \bibnamefont{{Sluys}}},
  \bibnamefont{and} \bibinfo{author}{\bibfnamefont{J.~G.~M.} \bibnamefont{{van
  Mier}}}, \bibinfo{journal}{Enginneering Fracture Mechanics}
  \textbf{\bibinfo{volume}{69}}, \bibinfo{pages}{1959}
  (\bibinfo{year}{2002}{\natexlab{a}}).

\bibitem[{\citenamefont{{Benz} and {Asphaug}}(1994)}]{Benz-e-Asphaug-1994}
\bibinfo{author}{\bibfnamefont{W.}~\bibnamefont{{Benz}}} \bibnamefont{and}
  \bibinfo{author}{\bibfnamefont{E.}~\bibnamefont{{Asphaug}}},
  \bibinfo{journal}{Icarus} \textbf{\bibinfo{volume}{107}}, \bibinfo{pages}{98}
  (\bibinfo{year}{1994}).

\bibitem[{\citenamefont{{Bouchaud} et~al.}(1993)\citenamefont{{Bouchaud},
  {Bouchaud}, {Lapasset}, and {Plan\`{e}s}}}]{Bouchaud-etal-1993}
\bibinfo{author}{\bibfnamefont{J.~P.} \bibnamefont{{Bouchaud}}},
  \bibinfo{author}{\bibfnamefont{E.}~\bibnamefont{{Bouchaud}}},
  \bibinfo{author}{\bibfnamefont{G.}~\bibnamefont{{Lapasset}}},
  \bibnamefont{and}
  \bibinfo{author}{\bibfnamefont{J.}~\bibnamefont{{Plan\`{e}s}}},
  \bibinfo{journal}{\prl} \textbf{\bibinfo{volume}{71}}, \bibinfo{pages}{2240}
  (\bibinfo{year}{1993}).

\bibitem[{\citenamefont{{Galybin} and {Dyskin}}(2004)}]{Galybin-e-Dyskin-2004}
\bibinfo{author}{\bibfnamefont{A.~N.} \bibnamefont{{Galybin}}}
  \bibnamefont{and} \bibinfo{author}{\bibfnamefont{A.~V.}
  \bibnamefont{{Dyskin}}}, \bibinfo{journal}{Int. J. Fracture}
  \textbf{\bibinfo{volume}{128}}, \bibinfo{pages}{95} (\bibinfo{year}{2004}).

\bibitem[{\citenamefont{{Paolicchi} et~al.}(1996)\citenamefont{{Paolicchi},
  {Verlicchi}, and {Cellino}}}]{Paolicchi-etal-1996}
\bibinfo{author}{\bibfnamefont{P.}~\bibnamefont{{Paolicchi}}},
  \bibinfo{author}{\bibfnamefont{A.}~\bibnamefont{{Verlicchi}}},
  \bibnamefont{and}
  \bibinfo{author}{\bibfnamefont{A.}~\bibnamefont{{Cellino}}},
  \bibinfo{journal}{Icarus} \textbf{\bibinfo{volume}{121}},
  \bibinfo{pages}{126} (\bibinfo{year}{1996}).

\bibitem[{\citenamefont{{Chang}
  et~al.}(2002{\natexlab{b}})\citenamefont{{Chang}, {Wang}, {Sluys}, and {van
  Mier}}}]{Chang-etal-2002a}
\bibinfo{author}{\bibfnamefont{C.~S.} \bibnamefont{{Chang}}},
  \bibinfo{author}{\bibfnamefont{T.~K.} \bibnamefont{{Wang}}},
  \bibinfo{author}{\bibfnamefont{L.~J.} \bibnamefont{{Sluys}}},
  \bibnamefont{and} \bibinfo{author}{\bibfnamefont{J.~G.~M.} \bibnamefont{{van
  Mier}}}, \bibinfo{journal}{Enginneering Fracture Mechanics}
  \textbf{\bibinfo{volume}{69}}, \bibinfo{pages}{1941}
  (\bibinfo{year}{2002}{\natexlab{b}}).

\bibitem[{\citenamefont{{Kun} and {Herrmann}}(1999)}]{Kun1999}
\bibinfo{author}{\bibfnamefont{F.}~\bibnamefont{{Kun}}} \bibnamefont{and}
  \bibinfo{author}{\bibfnamefont{H.~J.} \bibnamefont{{Herrmann}}},
  \bibinfo{journal}{\pre} \textbf{\bibinfo{volume}{59}}, \bibinfo{pages}{2623}
  (\bibinfo{year}{1999}).

\bibitem[{\citenamefont{Longaretti}(1989)}]{Longaretti1996}
\bibinfo{author}{\bibfnamefont{P.-Y.} \bibnamefont{Longaretti}},
  \bibinfo{journal}{Icarus} \textbf{\bibinfo{volume}{81}}, \bibinfo{pages}{51}
  (\bibinfo{year}{1989}).

\bibitem[{\citenamefont{{Spahn} et~al.}(2004)\citenamefont{{Spahn}, {Albers},
  {Sremcevic}, and {Thornton}}}]{Spahn-etal-2004}
\bibinfo{author}{\bibfnamefont{F.}~\bibnamefont{{Spahn}}},
  \bibinfo{author}{\bibfnamefont{N.}~\bibnamefont{{Albers}}},
  \bibinfo{author}{\bibfnamefont{M.}~\bibnamefont{{Sremcevic}}},
  \bibnamefont{and}
  \bibinfo{author}{\bibfnamefont{C.}~\bibnamefont{{Thornton}}},
  \bibinfo{journal}{Europhys. Lett.} \textbf{\bibinfo{volume}{67}},
  \bibinfo{pages}{545} (\bibinfo{year}{2004}).

\bibitem[{\citenamefont{Srem{\v c}evi{\'c} et~al.}(2007)\citenamefont{Srem{\v
  c}evi{\'c}, Schmidt, Salo, Sei{\ss\ }, Spahn, and Albers}}]{Sremcevic2007}
\bibinfo{author}{\bibfnamefont{M.}~\bibnamefont{Srem{\v c}evi{\'c}}},
  \bibinfo{author}{\bibfnamefont{J.}~\bibnamefont{Schmidt}},
  \bibinfo{author}{\bibfnamefont{H.}~\bibnamefont{Salo}},
  \bibinfo{author}{\bibfnamefont{M.}~\bibnamefont{Sei{\ss\ }}},
  \bibinfo{author}{\bibfnamefont{F.}~\bibnamefont{Spahn}}, \bibnamefont{and}
  \bibinfo{author}{\bibfnamefont{N.}~\bibnamefont{Albers}},
  \bibinfo{journal}{Nature} \textbf{\bibinfo{volume}{449}},
  \bibinfo{pages}{1019} (\bibinfo{year}{2007}).

\bibitem[{\citenamefont{Brilliantov et~al.}(2007)\citenamefont{Brilliantov,
  Albers, Spahn, and Poeschel}}]{Brilliantov2007}
\bibinfo{author}{\bibfnamefont{N.}~\bibnamefont{Brilliantov}},
  \bibinfo{author}{\bibfnamefont{N.}~\bibnamefont{Albers}},
  \bibinfo{author}{\bibfnamefont{F.}~\bibnamefont{Spahn}}, \bibnamefont{and}
  \bibinfo{author}{\bibfnamefont{T.}~\bibnamefont{Poeschel}},
  \bibinfo{journal}{Phys. Rev. E} \textbf{\bibinfo{volume}{76}},
  \bibinfo{pages}{051302} (\bibinfo{year}{2007}).

\bibitem[{\citenamefont{Krapivsky et~al.}(2010)\citenamefont{Krapivsky, Redner,
  and Ben-Naim}}]{Krap2010}
\bibinfo{author}{\bibfnamefont{P.~L.} \bibnamefont{Krapivsky}},
  \bibinfo{author}{\bibfnamefont{S.}~\bibnamefont{Redner}}, \bibnamefont{and}
  \bibinfo{author}{\bibfnamefont{E.}~\bibnamefont{Ben-Naim}},
  \emph{\bibinfo{title}{A Kinetic View of Statistical Physics}}
  (\bibinfo{publisher}{Cambridge University Press}, \bibinfo{year}{2010}).

\end{thebibliography}

\end{document}